# A Unified Framework for the Ergodic Capacity of Spectrum Sharing Cognitive Radio Systems

Lokman Sboui, Zouheir Rezki and Mohamed-Slim Alouini
Electrical Engineering Program
Computer, Electrical and, Mathematical Sciences and Engineering (CEMSE) Division
King Abdullah University of Science and Technology (KAUST)
Thuwal, Makkah Province, Saudi Arabia
{lokman.sboui,zouheir.rezki,slim.alouini}@kaust.edu.sa

*Abstract*—We consider a spectrum sharing communication scenario in which a primary and a secondary users are communicating, simultaneously, with their respective destinations using the same frequency carrier. Both optimal power profile and ergodic capacity are derived for fading channels, under an average transmit power and an instantaneous interference outage constraints. Unlike previous studies, we assume that the secondary user has a noisy version of the cross link and the secondary link Channel State Information (CSI). After deriving the capacity in this case, we provide an ergodic capacity generalization, through a unified expression, that encompasses several previously studied spectrum sharing settings. In addition, we provide an asymptotic capacity analysis at high and low signal-to-noise ratio (SNR). Numerical results, applied for independent Rayleigh fading channels, show that at low SNR regime, only the secondary channel estimation matters with no effect of the cross link on the capacity; whereas at high SNR regime, the capacity is rather driven by the cross link CSI. Furthermore, a practical on-off power allocation scheme is proposed and is shown, through numerical results, to achieve the full capacity at high and low SNR regimes and suboptimal rates in the medium SNR regime.

*Index Terms*—Underlay cognitive radio, Spectrum sharing, Ergodic capacity, Imperfect CSI, On-off scheme.

## I. Introduction

Traditional spectrum allocation is presenting a huge inefficiency since most of cellular networks bands are overloaded while other bands are not efficiently exploited and the usage peak does not exceed 15% [1]. As a solution for this issue, the cognitive radio (CR) concept was introduced, in 1999, by Mitola and Maguire [2]. CR systems are based on "smart systems" that sense for spectrum holes and share already-licensed frequencies and adapt communication parameters (frequency, time, coding, power, etc.) depending on licensed user's situations and channel state in a way of preserving primary communication quality.

A crucial measure of the CR performance is the capacity of the secondary communication. Fundamental capacity limits of CR system was studied along with transmission techniques in [3]. From an industrial perspective, an overview of standards based on CR was presented in [4]. Meanwhile, a general study of CR networks focusing on practical layers implementation was presented in [5].

Depending on the available knowledge of the CR system, three CR modes were introduced [3]: underlay, overlay, and interweave. In underlay mode, the CR system has "some" knowledge of the primary channel state and channel gain and adapts its transmission below a tolerated interference constraint [6]–[8]. In overlay mode, the CR system knows primary message, codebook and channel gain, and transmits at any power since noncognitive message is relayed (retransmitted by the CR system) [9], [10]. In the interweave mode, the CR system senses primary spectrum holes (in frequency, space and time), and establishes opportunistic connections [11], [12].

In underlay mode, a study of the spectrum sharing approach involving unlicensed users was presented in [13] providing methods and techniques to avoid harmful interference. Moreover, the spectrum propriety rights and related regulations were highlighted in [14]. Nevertheless, the Quality of Service (QoS) of the primary user has to be maintained at a good level by imposing an interference constraint on the secondary transmitter [15]. That is, the interference power measured by the primary user should not exceed a certain threshold called interference temperature [16]. In previous works related to spectrum sharing, the fundamental capacity limit of AWGN channels was studied in [6]. In addition, a general fading capacity was discussed either with perfect channel state information (CSI) for both links in [7], [17] or with perfect secondary link channel state information (SL-CSI) and estimated cross link CSI (CL-CSI) in [18]. Meanwhile, the scenario of perfect SL-CSI and estimated CL-CSI was analyzed in [8]. Nevertheless, the capacity of both estimated links was studied in [19] with an interference constraint and in [20] with power constrains. The effect of the spectrum sharing on the primary capacity was studied in [21].

In this paper, we study the spectrum sharing CR underlay mode in which a secondary unlicensed user is allowed to share opportunistically the spectrum with a primary licensed user, under some constraints that aim at maintaining a good QoS level of the primary communication. As a first step, we derive the ergodic capacity of a spectrum sharing system under imperfect CSI at the secondary transmitter (CSI-T). Since in practice, perfect CSI is hard to obtain, our newly reported results are useful to understand the fundamental performance of cognitive radio systems in practical settings and how channel uncertainty would impact performances. Then, we build on our general framework and provide generic expressions of the optimal power profile and the ergodic capacity, for different level of CSI-T. When properly applied, the later results retrieve previously presented results in e.g., [7], [18], [22], [23] as special cases. Then, we study the effect of estimation error on the performance of the CR system at low and high SNR. Finally, we propose a practical on-off scheme that is shown to be capacity-achieving, through numerical simulations at low and high SNR regimes with low complexity. Our contribution in this paper is summarized in the following points:

- Providing the expressions of the optimal power and ergodic capacity of an estimated CSI for both SL and CL spectrum sharing system with average power and outage interference constraints. These expression could be presented in closed form once the fading distribution is known.
- Establishing a framework that encloses different level of knowledge of SL-CSI and CL-CSI as well as different fading distribution.
- Performing an asymptotic analysis at high and low SNR.
- Studying the performance of the on-off scheme as an alternative to the optimal power scheme which is relatively complex.

The rest of the paper is organized as follows. The system model is presented in Section II. The optimal power profile along with the ergodic capacity of estimated SL and CL-CSI are derived in Section III. A unified expression of the capacity along with simulations of some previously studied cases are presented in Section IV. The capacity behavior at high and low SNR regime is studied in Section V. An On-off scheme is then introduced in Section VI. Finally, the conclusion of the paper is presented in Section VII.

## II. System Model

We consider a spectrum sharing scenario as shown in Fig. 1. Both secondary and primary users (SU and PU) are performing a point to point communication using the same narrow-band frequency. We assume that the secondary user can decode the primary transmission so that there is no interference caused by the primary user on the secondary communication as in [24], where the channel gains are perfectly known and in [8] where the authors adopt either the average power constraint or peak interference constraint in separate problems. In our framework, we consider the two following links: the secondary link (SL) between the secondary transmitter (SU_Tx) and the secondary receiver (SU_Rx) and the cross link (CL) between the secondary transmitter (SU_Tx) and the primary receiver (PU_Rx). Each link is characterized by a complex channel gain, noted $h_s$ for the SL and $h_p$ for the CL and assumed to be independent. We, also, assume that the (SU_Tx) performs a minimum mean square error (MMSE) estimation of the channel gain and we adopt, for each link, the following model:

$$h_i = \hat{h}_i + \tilde{h}_i, \qquad i = \{s, p\}, \qquad (1)$$

where $\hat{h}_s$ and $\hat{h}_p$ are the secondary and the cross links channel MMSE estimations, respectively, and $\tilde{h}_s$ and $\tilde{h}_p$ are their respective estimation errors. Note that the MMSE model ensures that $\hat{h}_i$ and $\tilde{h}_i$ are uncorrelated for $i = \{s, p\}$. We note by $\alpha_s$ and $\alpha_p$ the two variances of $\tilde{h}_s$ and $\tilde{h}_p$, respectively. We assume that in our CR system, the SU knows both $\hat{h}_s$ and $\hat{h}_p$ along with the statistics and the distribution functions. Hence, the SU_Tx is able to optimize transmission power that achieves the capacity with respect to an average power constraint in a way that maintains an acceptable primary QoS. It should be pointed out that the channel MMSE estimation model in (1) has been used in many previous works either in channel estimation contexts such as in [25] or in CR frameworks such as in [24] and [8]. More specifically, in our model we adopt two constraints on the secondary transmission power: an average transmit power constraint at the SU_Tx [26], and an interference outage constraint characterizing how tolerant the PU_Rx is toward violating an interference threshold by the SU. Meanwhile, the average transmit power constraint is adopted since the SU is considered to be a mobile device powered by a limited source such as a battery. On the other hand, the interference constraint, unlike the average power constraint, does not average over all channel gain values and it presents a high dynamicity. Thus the power should be, instantaneously, adapted as the channel is varying.

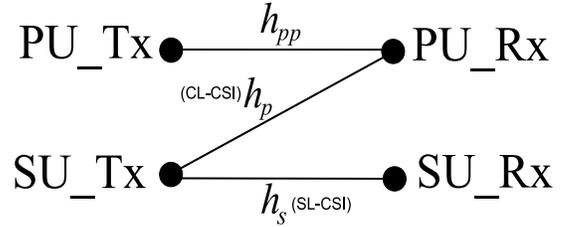

Figure 1. Spectrum sharing system model.

Let $H$ be a set of channel gains on which the averaging is performed to obtain the ergodic capacity. For example, if both secondary and cross links present imperfect CSI, $H = \{(h_s, \hat{h}_s, \hat{h}_p)\}$. For a given value of the transmission power $P$, an achievable rate of the SU_Tx derived by averaging over all states of the channels $H$ is:

$$R(P(\hat{h}_s, \hat{h}_p)) = \mathbb{E}_H \left[ \log(1 + P(\hat{h}_s, \hat{h}_p) \times |h_s|^2) \right], \qquad (2)$$

and the corresponding ergodic capacity is given by

$$C = \max_{P(\hat{h}_s, \hat{h}_p) \in \Pi} R(P(\hat{h}_s, \hat{h}_p)), \qquad (3)$$

where $\mathbb{E}[\cdot]$ is the expectation function and $\Pi$ is the set of transmission power respecting the adopted constraints. We note by $P^*$ the power that maximize $R$. $P^*$ is called the optimal power and defined as a function of $H$.

Note that in [21], the authors study the spectrum sharing system from the primary capacity point of view. Their framework describes the effect of sharing the spectrum on the primary throughput. Whereas in our work we focus on secondary capacity.

## III. Ergodic Capacity with Imperfect CSI

### A. Problem Formulation

In this problem, we assume that the SU_Tx is aware of $\hat{h}_s$ and $\hat{h}_p$ and thus, the power is adapted through the estimation of the SL and CL channel gains, i.e., $P = P(\hat{h}_s, \hat{h}_p)$. In this case, $H$ in (2) is defined as $H = \{(h_s, \hat{h}_s, \hat{h}_p)\}$. The ergodic capacity is then determined by [27]–[29] as

$$C = \max_{P(\hat{h}_s, \hat{h}_p) \geq 0} \mathbb{E}_{h_s, \hat{h}_s, \hat{h}_p}[\log(1 + P(\hat{h}_s, \hat{h}_p) \times |h_s|^2)]. \quad (4)$$

In our model, the power $P(\hat{h}_s, \hat{h}_p)$ is constrained by

$$\mathbb{E}_{\hat{h}_s, \hat{h}_p}[P(\hat{h}_s, \hat{h}_p)] \leq P_{avg} \quad (5)$$

and $\quad \text{Prob}\left\{P(\hat{h}_s, \hat{h}_p)|h_p|^2 \geq I_{peak}\big|\hat{h}_p\right\} \leq \varepsilon, \quad (6)$

respectively. Beside the average transmit power constraint in (5), we adopt the constraint (6) in order to preserve the primary communication by respecting an outage threshold, noted $\varepsilon$, along with the dynamic variation of the primary channel gain $h_p$ which was adopted in [19]. However, we perform a conditioning on the estimated channel gain $\hat{h}_p$. This constraint may be interpreted as an instantaneous outage constraint. Thus, the power is instantaneously adapted as $\hat{h}_p$ varies.

The constraint in (6) can be written equivalently as (See Appendix I)

$$P(\hat{h}_s, \hat{h}_p) \leq P_{I_{peak}, \varepsilon}(\hat{h}_p), \quad (7)$$

where $P_{I_{peak}, \varepsilon}(\hat{h}_p) = \frac{I_{peak}}{F^{-1}_{|h_p|^2|\hat{h}_p}(1-\varepsilon, \hat{h}_p)}$, and $F^{-1}_{|h_p|^2|\hat{h}_p}(\cdot, \hat{h}_p)$ is the inverse of the cumulative density function (c.d.f.) of $|h_p|^2$ given $\hat{h}_p$. As such, the ergodic capacity can be derived by solving the following maximization problem

$$C = \max_{P(\hat{h}_s, \hat{h}_p) \geq 0} \mathbb{E}_{h_s, \hat{h}_s, \hat{h}_p}\left[\log\left(1 + P(\hat{h}_s, \hat{h}_p) \times |h_s|^2\right)\right] \quad (8)$$

$$\text{subject to} \quad \mathbb{E}_{\hat{h}_s, \hat{h}_p}\left[P(\hat{h}_s, \hat{h}_p)\right] \leq P_{avg}, \quad (9)$$

$$\text{and} \quad P(\hat{h}_s, \hat{h}_p) \leq P_{I_{peak}, \varepsilon}(\hat{h}_p). \quad (10)$$

Note that in previous works, the authors in [19] adopt the interference constraints with no conditioning on the estimated value of the channel gain $\hat{g}_0$. Moreover, the study of the cognitive capacity is performed as a function of the interference threshold $I_{th}$, as well as in [8]. In our work the capacity is studied as a function of the cognitive average power $P_{avg}$ by adopting two constraints: average power and peak interference constraint in which perform a conditioning over the estimated value of the channel gain in equation (6). In addition, in [20], the authors adopt an average interference power in (8), where in our framework, we adopt an instantaneous probabilistic interference constraint which is more realistic in a spectrum sharing scenario. Whereas, in [22], the authors adopt either the average power constraint or peak interference constraint in separate problems. In our framework, both constraints are considered.

### B. General Solution

We use the Lagrangian method to find $C$, which gives a necessary and sufficient optimality condition due to the concavity of the problem [30] (see Appendix II).

$$\int_{|h_s|^2} \frac{\gamma}{1 + P(\hat{h}_s, \hat{h}_p) \times \gamma} f_{|h_s|^2|\hat{h}_s}(\gamma, \hat{h}_s) \, d\gamma = \lambda + \mu, \ P \geq 0, \quad (11)$$

where $f_{|h_s|^2|\hat{h}_s}(\cdot, \hat{h}_s)$ is the probability density function (p.d.f.) of $|h_s|^2$ conditioned on $\hat{h}_s$ and $\lambda$ and $\mu$ are the Lagrangian multipliers corresponding to constraints (5) and (6), respectively. Note that even without cognitive constraint, (11) has been analyzed in [28] and no closed form solution has been found. We therefore expect the problem to be more involved when the cognitive constraint is introduced. Nevertheless, we have the following result:

**Theorem 1**

For the spectrum sharing channel model described by (1), the optimal power profile of the secondary link under constraints (5) and (6), is given by

$P_{opt}(\hat{h}_s, \hat{h}_p) =$

$$\begin{cases} \min\left\{P_{P_{avg}}(\hat{h}_s), P_{I_{peak}, \varepsilon}(\hat{h}_p)\right\} & \text{if } P_{avg} \leq \mathbb{E}_{\hat{h}_s, \hat{h}_p}\left[P_{I_{peak}, \varepsilon}(\hat{h}_p)\right] \\ P_{I_{peak}, \varepsilon}(\hat{h}_p), & \text{otherwise,} \end{cases} \quad (12)$$

where $P_{P_{avg}}(\hat{h}_s)$ is the solution of (11) and the resulting ergodic capacity of the secondary link is given by

$$C = \int_{\hat{h}_p} \left(\int_{\hat{h}_s} \left(\int_{|h_s|^2} \log\left(1 + P_{opt}(\delta, \nu)\gamma\right) f_{|h_s|^2|\hat{h}_s}(\gamma, \delta) \, d\gamma\right) \right. \\ \left. \times f_{\hat{h}_s}(\delta) \, d\delta\right) f_{\hat{h}_p}(\nu) \, d\nu. \quad (13)$$

*Proof:* The proof is presented in the Appendix. ∎

Note that once the fading type is fixed, the expression of the capacity can be computed explicitly. This expression may be seen as a generalization of related problems as follows:

- CSI knowledge; each link presents three levels of knowledge: perfectly known ($\alpha_i = 0$), imperfect ($\alpha_i \in ]0, 1[$) and unknown ($\alpha_i = 1$) with $i = \{p, s\}$.
- Fading type; for simulation we adopt a Rayleigh fading, however any continuous PDF fading could be implemented in (13) to derive the related capacity.

### C. Example of Rayleigh Fading

Numerical results are presented for Rayleigh fading in which, for $\imath = \{p, s\}$, we have $h_i \sim \mathcal{CN}(0, 1)$, complex normal distribution with zero mean and variance equal to 1. Moreover, considering (1), channel gain estimations and errors are given by $\hat{h}_i \sim \mathcal{CN}(0, 1 - \alpha_i)$ and $\tilde{h}_i \sim \mathcal{CN}(0, \alpha_i)$, respectively, and $\alpha_i$ is the error variance of the MMSE estimator of $h_i$ for $i = \{p, s\}$. In this scenario, $\hat{h}_i$ and $\tilde{h}_i$ are independent by the propriety of the MMSE estimation. Note that in the case of Rayleigh fading channels, the power depends only on the modulus of the channel gains since the phase is uniform. The PDF are given by

$$f_{|h_i|^2}(\gamma) = e^{-\gamma} \ ; \ f_{|\hat{h}_i|^2}(u) = \frac{1}{1 - \alpha_i} e^{-\frac{u}{1-\alpha_i}} \ ; \ \gamma, u \geq 0,$$

$$f_{|h_i|^2|\hat{h}_i}(\gamma, \hat{h}_i) = \frac{1}{\alpha_i} e^{-\frac{\gamma + |\hat{h}_i|^2}{\alpha_i}} I_0\left(\frac{2}{\alpha_i}\sqrt{|\hat{h}_i|^2\gamma}\right) \ ; \ \gamma \geq 0, \text{ for } i = \{p, s\}$$

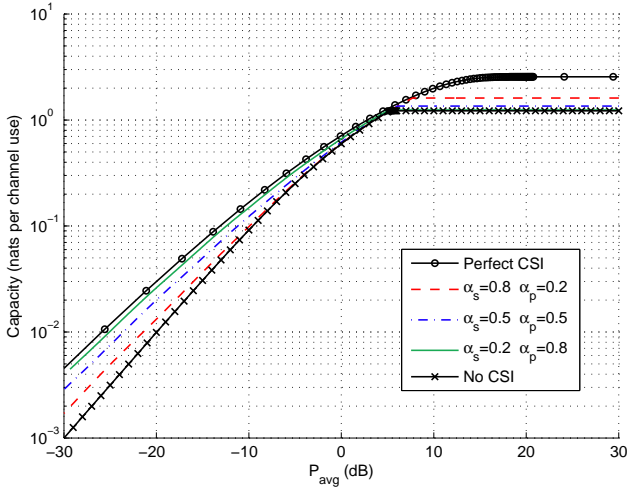

Figure 2. CR optimal power profile for estimated CSI-T in logarithmic scale.

where $I_0(\cdot)$ is the modified Bessel function of the first kind [31]. In addition, as $h_s$ and $\hat{h}_p$ are independent we have

$$f_{|h_s|^2|\hat{h}_p}(\gamma) = f_{|h_s|^2}(\gamma) = e^{-\gamma} \text{ and } f_{|h_p|^2|\hat{h}_s}(\delta) = f_{|h_p|^2}(\delta) = e^{-\delta}.$$

Simulation results are shown in Fig. 2 with $I_{peak} = 10$ dB and $\varepsilon = 0.05$ (5%). We notice that, at low SNR, the capacity is similar to the capacity of single-user communication with estimated CSI-T [28] and depends only on the knowledge level of the SL-CSI. However at high SNR, the capacity is capped by a constant as $P_{avg}$ increases. This asymptotic constant of the capacity at high values of $P_{avg}$ varies depending on the available CL-CSI, from 2.55 nats per channel use (npcu) for perfect CSI-T ($\alpha = 0$) to 1.36 npcu for estimated CSI-T with $\alpha_p = 0.5$. This means that above some value of $P_{avg}$, say $P^*_{avg}$, an increase of the average power does not increase the capacity. Hence, there is no need to exceed this $P^*_{avg}$. In order to find $P^*_{avg}$, we proceed by formulating the dual problem. The resulting power solution presents two regimes depending on the variation of $P_{avg}$ as follows

- If $P_{avg} \leqslant \mathbb{E}_{\hat{h}_s,\hat{h}_p}\left[P_{I_{peak},\varepsilon}(\hat{h}_p)\right]$, then the capacity profile is similar to the non cognitive constrained scenario profile in agreement with [32].
- If $P_{avg} > \mathbb{E}_{\hat{h}_s,\hat{h}_p}\left[P_{I_{peak},\varepsilon}(\hat{h}_p)\right]$ then the capacity present a saturation as shown in [18]. In fact, by adopting the power described in (12) the resulting capacity does not depend on $P_{avg}$ any more which explains the capacity saturation.

Hence we can conclude that

$$P^*_{avg} = \mathbb{E}_{\hat{h}_s,\hat{h}_p}\left[P_{I_{peak},\varepsilon}(\hat{h}_p)\right]. \quad (14)$$

Note that in Section V, a broader study of the capacity at low and high SNR is presented.

## IV. UNIFIED CAPACITY AND APPLICATION ON PREVIOUS WORKS

The previous results of imperfect CSI-T for both CL and SL as well as results in [33] and [18] give an insightful generalization of the expression of the capacity under any CL and SL knowledge level. In this Section, a unified expression of the optimal power and the capacity for various CSI-T knowledge is presented. This generalizes several previous results i.e., [7], [18], [33]. Then we present, two special cases corresponding to either perfect SL-CSI or perfect CL-CSI scenarios.

### A. General Expression for the Optimal Power

From equation (12), we notice that the SU_Tx optimal power of the cognitive radio scenario is composed of two parts

- The first part depends on $h_s$ and is related to the average power constraint and is a function of $P_{avg}$, noted $P_{P_{avg}}(h_s)$.
- The second part depends on $h_p$ and is related to the peak power constraint and is a function of $I_{peak}$ noted, $P_{I_{peak}}(h_p)$.

Depending on the knowledge of the SL-CSI and the CL-CSI and on the adopted parameters ($\varepsilon$, $\alpha_s$, $\alpha_p$), the expressions of $P_{P_{avg}}(h_s)$ and $P_{I_{peak}}(h_p)$ are computed differently for each scenario. Considering all combinations of SL-CSI and CL-CSI knowledge, the general expression of the optimal power $P_{opt}$ is given by

$$P_{opt} = \begin{cases} \min\left\{P_{P_{avg}}(h_s), P_{I_{peak}}(h_p)\right\}, & \text{if } P_{avg} \leqslant \mathbb{E}_{h_s,h_p}\left[P_{I_{peak}}(h_p)\right] \\ P_{I_{peak}}(h_p), & \text{otherwise,} \end{cases} \quad (15)$$

where $P_{P_{avg}}$ and $P_{I_{peak}}$ are picked from Table I. In this table, we find, again, that the expression of $P_{P_{avg}}$ is related to the knowledge of the SL-CSI whereas $P_{I_{peak}}$ depends on the CL-CSI. Moreover, in order to preserve generality, the Lagrangian multiplier $\lambda$ in the expression of $P_{P_{avg}}$ is not explicitly presented since its value depends on the adopted fading channel and the chosen $P_{I_{peak}}(h_p)$. Recall that $\lambda$ is computed by solving

$$\mathbb{E}_{H_0}\left[\min\left\{P_{P_{avg}}(h_s), P_{I_{peak}}(h_p)\right\}\right] = P_{avg}, \quad (16)$$

where $H_0$ is the set of channel gains given in Table II.

### B. General Expression for Capacity

Once the optimal power is computed, the ergodic capacity is easily expressed as

$$C = \mathbb{E}_{H_1}\left[\log\left(1 + P_{opt} \times |h_s|^2\right)\right], \quad (17)$$

where $H_1$ is the set of channel gains given in Table II. Hence, for fixed $P_{avg}, I_{peak}$ and ($\varepsilon, \alpha_s, \alpha_p$), the capacity is computed by classic numerical integrations. This unified capacity expression encloses all related scenarios, adopting the same constraints, but with various knowledge of CSI and/or channel fading distributions. In practical simulations, the difficulty appears in finding the optimal $\lambda$ that satisfies (16). For this reason, to provide numerical results, we start from a value of $\lambda$, then we determine the associated power. Finally, the resulting capacity $C$ and average power $P_{avg}$ are given by integrating the computed power over the channel gains as indicated in (17). Algorithm 1 describes how to plot the capacity as function of $P_{avg}$.

## Algorithm 1 Capacity computing algorithm

1: **for** fixed $\alpha_s \in ]0, 1[$ and $\alpha_p \in ]0, 1[$ **do**
2:   **for** $\lambda \geq 0$ **do**
3:     For all $|\hat{h}_s|^2 \geq 0, |\hat{h}_p|^2 \geq 0$ Compute $P_{avg}$ from (16)
4:     **for** $|\hat{h}_s|^2 \geq 0$ **do**
5:       Compute $P_{P_{avg}}(|\hat{h}_s|^2)$ from Table I
6:       **for** $|\hat{h}_p|^2 \geq 0$ **do**
7:         Compute $P_{I_{peak}}(|\hat{h}_p|^2)$ from Table I
8:         Determine $P_{opt}$ from (15) using the derived value of $P_{avg}$
9:       **end for**
10:     **end for**
11:     Compute $C$ such as $C = \mathbb{E}_{h_s, \hat{h}_s, \hat{h}_p}[\log(1 + P_{opt} \times |h_s|^2)]$
12:   **end for**
13:   Plot $C$ as a function of $P_{avg}$
14: **end for**

### C. Special Cases

In this section, two examples of different knowledge of CSI-T are discussed as illustration of the framework that encloses unified power and capacity expressions presented in (15) and (17).

#### 1) Perfect SL-CSI and Estimated CL-CSI:

*a) Problem Formulation:* This scenario describes the situation where the SU_Tx is perfectly aware of the secondary link state (SL-CSI) but has a noisy version of $\hat{h}_p$; the cross link state (CL-CSI). This problem is formulated in the following maximization problem

$$C = \max_{P(h_s, \hat{h}_p) \geq 0} \mathbb{E}_{h_s, \hat{h}_p}[\log(1 + P(h_s, \hat{h}_p) \times |h_s|^2)], \quad (18)$$

$$\text{subject to} \quad \mathbb{E}_{h_s, \hat{h}_p}[P(h_s, \hat{h}_p)] \leq P_{avg}, \quad (19)$$

$$\text{and} \quad \text{Prob}\left\{P(h_s, \hat{h}_p) \times |h_p|^2 \geq I_{peak} | \hat{h}_p\right\} \leq \varepsilon. \quad (20)$$

*b) General Solution:* The optimal power that maximizes the capacity is chosen, according to (15), from Table I as

$$P(\hat{h}_s, \hat{h}_p) = \begin{cases} \min\{P_{P_{avg}}(h_s), P_{I_{peak}}(\hat{h}_p)\}, & \text{if } P_{avg} \leq \mathbb{E}_{h_s, \hat{h}_p}\left[P_{I_{peak}}(\hat{h}_p)\right] \\ P_{I_{peak}}(\hat{h}_p), & \text{otherwise,} \end{cases} \quad (21)$$

where $P_{P_{avg}}(h_s) = \left[\frac{1}{\lambda} - \frac{1}{|h_s|^2}\right]^+$ and $P_{I_{peak}}(\hat{h}_p) = \frac{I_{peak}}{F^{-1}_{|h_p|^2|\hat{h}_p}(1 - \varepsilon, \hat{h}_p)}$. (22)

In this scenario, $\lambda$ is determined as a solution of

$$\mathbb{E}_{H_0 = \{h_s, \hat{h}_p\}}\left[\min\left\{\left[\frac{1}{\lambda} - \frac{1}{|h_s|^2}\right]^+, \frac{I_{peak}}{F^{-1}_{|h_p|^2|\hat{h}_p}(1 - \varepsilon, \hat{h}_p)}\right\}\right] = P_{avg}. \quad (23)$$

Hence, the ergodic capacity is given by

$$C = \begin{cases} \int_{\hat{h}_p} \left(\int_{|h_s|^2} (\log\left(1 + \min\left\{\left[\frac{1}{\lambda} - \frac{1}{\gamma}\right]^+, \frac{I_{peak}}{F^{-1}_{|h_p|^2|\hat{h}_p}(1-\varepsilon, \nu)}\right\}\gamma\right) \\ \times f_{h_s}(\gamma)\right) d\gamma\right) f_{\hat{h}_p}(\nu) \, d\nu, \\ \quad \text{if } P_{avg} \leq \mathbb{E}_{\hat{h}_s, h_p}\left[P_{I_{peak}}(h_p)\right] \\ \\ \int_{\hat{h}_p} \left(\int_{|h_s|^2} \left(\log(1 + \frac{I_{peak}}{F^{-1}_{|h_p|^2|\hat{h}_p}(1-\varepsilon, \nu)}\gamma\right) f_{h_s}(\gamma) \, d\gamma\right) f_{\hat{h}_p}(\nu) \, d\nu, \\ \quad \text{otherwise} \end{cases} \quad (24)$$

Note that, this capacity expression is, again, in agreement with [18], [34].

#### 2) Estimated SL-CSI and Perfect CL-CSI:

*a) Problem Formulation:* This scenario describes a situation, for instance, whereas feedback link from the SU_Rx is of low capacity, where the feedback link from PU_Rx is of sufficient capacity that SU_Tx is able to track the channel perfectly.

Note that, to the best of our knowledge, this problem has not been addressed before. In this case the maximization problem is

$$C = \max_{P(\hat{h}_s, h_p) \geq 0} \mathbb{E}_{h_s, \hat{h}_s, h_p}[\log(1 + P(\hat{h}_s, h_p) \times |h_s|^2)], \quad (25)$$

$$\text{subject to} \quad \mathbb{E}_{\hat{h}_s, h_p}[P(\hat{h}_s, h_p)] \leq P_{avg}, \quad (26)$$

$$\text{and} \quad P(\hat{h}_s, h_p) \leq \frac{I_{peak}}{|h_p|^2}. \quad (27)$$

*b) General Solution:* The optimal power that maximizes the ergodic capacity is given as mentioned previously by

$$P(\hat{h}_s, h_p) = \begin{cases} \min\{P_{P_{avg}}(\hat{h}_s), P_{I_{peak}}(h_p)\}, & \text{if } P_{avg} \leq \mathbb{E}_{\hat{h}_s, h_p}\left[P_{I_{peak}}(h_p)\right] \\ P_{I_{peak}}(h_p), & \text{otherwise} \end{cases} \quad (28)$$

where $P_{P_{avg}} = \left[I^{-1}_{\hat{h}_s}(\lambda)\right]^+$ (from Table I) and $P_{I_{peak}}(h_p) = \frac{I_{peak}}{|h_p|^2}$. (29)

In (29), $\lambda$ is determined as a solution of

$$\mathbb{E}_{H_0 = \{\hat{h}_s, h_p\}}\left[\min\left\{\left[I^{-1}_{\hat{h}_s}(\lambda)\right]^+, \frac{I_{peak}}{|h_p|^2}\right\}\right] = P_{avg}. \quad (30)$$

Finally, the ergodic capacity is given by

$$C = \begin{cases} \int_{|h_p|^2} \left(\int_{\hat{h}_s} \left(\int_{|h_s|^2} (\log(1 + \min\{P_{P_{avg}}(\gamma), \frac{I_{peak}}{\nu}\} \\ \gamma) f_{|h_s|^2|\hat{h}_s}(\gamma, \delta)) \, d\gamma\right) f_{\hat{h}_s}(\delta) \, d\delta\right) f_{|h_p|^2}(\nu) \, d\nu \\ \quad \text{if } P_{avg} \leq \mathbb{E}_{\hat{h}_s, h_p}\left[P_{I_{peak}}(h_p)\right] \\ \\ \int_{|h_p|^2} \left(\int_{\hat{h}_s} \left(\int_{|h_s|^2} (\log(1 + \frac{I_{peak}}{\nu}\gamma) \\ f_{|h_s|^2|\hat{h}_s}(\gamma, \delta)) \, d\gamma\right) f_{\hat{h}_s}(\delta) \, d\delta\right) f_{|h_p|^2}(\nu) \, d\nu, \\ \quad \text{otherwise.} \end{cases} \quad (31)$$

## V. Asymptotic Analysis

In this section, we study the ergodic capacity behavior at low SNR ($P_{avg} \to 0$) and at high SNR ($P_{avg} \to \infty$) in order to gain some insights on the capacity behavior at these regimes.

### A. Low SNR Regime

*1) General Capacity Limit:* At fixed $I_{peak}$, when $P_{avg} \to 0$ so that $P_{avg} \ll I_{peak}$, the power allocation problem is only constrained by the average transmit power and the optimal power is the power that satisfies this constraint with equality. Hence, $\lim_{P_{avg}\to 0} P = P_{P_{avg}}$, where $P_{P_{avg}}$ is the power component related to the average power constraint and defined in Table I. Consequently, the capacity limit at low SNR is given by

$$\lim_{P_{avg}\to 0} C = \mathbb{E}_{H_3}\left[\log(1 + P_{P_{avg}} \times |h_s|^2)\right], \quad (32)$$

where $H_3$ is given by the first row of $H_1$ in Table II, since the capacity, at low SNR regime, is independent of the CL-CSI state. Hence, depending on the available SL-CSI knowledge, different reduced forms of (32) are given in Table III.

*2) Rayleigh Fading Capacity Limit:* We derive the expression of the Rayleigh fading capacity at low SNR according to (32) for all SL-CSI knowledge states. Derived results are considered as capacity asymptotes and are presented in Table III.

Note that, depending on the level of knowledge of the SL-CSI, the capacity expression may be reduced to a non-integral closed form (for $\alpha_s \in \{0, 1\}$). However, in estimated CSI-T $\alpha_s \in ]0, 1[$, the capacity is computed by relying on numerical integrations. In Fig. 2 the ergodic capacity is presented versus the average transmit power at the SU_Tx for different CL-CSI and SL-CSI estimation errors. The graph is presented in logarithmic scale to highlight capacity at low SNR. We notice that, at low SNR range ($\leq 0$ dB), our graph is close to a single-user communication developed in [28]. This result confirms that at low SNR the SU obtains the same capacity either with or without the existence of the PU. Hence the SU_Tx is not obliged to estimate the CL-CSI. Instead, the SL-CSI has to be well estimated in order to achieve a better capacity.

*3) Effect of the SL-CSI Estimation Error $\alpha_s$:* In order to study the effect of the SL-CSI estimation variance, $\alpha_s$, on the capacity limit at low SNR with a fixed $P_{avg}$, the variation of the capacity with $\alpha_s$ is plotted numerically in Fig. 3 for $P_{avg} = -25$ dB. We notice that the capacity is decreasing as $\alpha_s$ increases in an almost linear slope. This result is again in full agreement with recent results where it has been established that the capacity varies essentially as $(1 - \alpha_s) \times P_{avg} \log \frac{1}{P_{avg}}$ at low-SNR [35].

### B. High SNR Regime

*1) General Capacity Limit:* At fixed $I_{peak}$, when $P_{avg} \to +\infty$ the power allocation problem is only constrained by the interference constraint and the optimal power is the power that satisfies this constraint with equality. Hence $\lim_{P_{avg}\to+\infty} P = P_{I_{peak}}(h_p)$ where $P_{I_{peak}}(h_p)$ is the power component related to

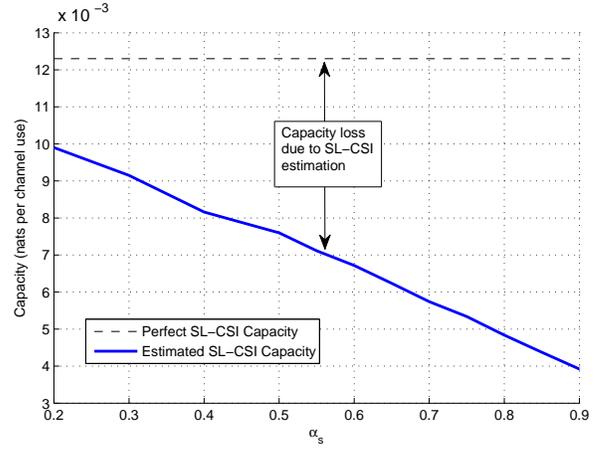

Figure 3. SL-CSI estimation error effect on the capacity at low SNR with $P_{avg} = -25$ dB.

the interference constraint and defined in Table I. Consequently, the capacity limit at high SNR is given by

$$\lim_{P_{avg}\to+\infty} C = \mathbb{E}_{H_4}\left[\log(1 + P_{I_{peak}}(h_p) \times |h_s|^2)\right], \quad (33)$$

where $H_4$ is a set of channel gains on which the expectation will be performed, depending on different levels of knowledge of SL-CSI or CL-CSI. Since there is no effect of the knowledge of the SL-CSI on the capacity at high SNR regime, $H_4$ is given by the first column of Table II corresponding to $H_1$. Depending on the available CL-CSI knowledge, different reduced forms of (33) are given in Table IV.

*2) Rayleigh Fading Capacity Limit:* Similarly to the low SNR case, we find that the CL-CSI capacity expression may be reduced to a non-integral closed form (for $\alpha_p \in \{0, 1\}$). However, in estimated CSI-T $\alpha_p \in ]0, 1[$, the capacity is computed numerically. Capacity expression for Rayleigh fading are presented in Table IV in this case.

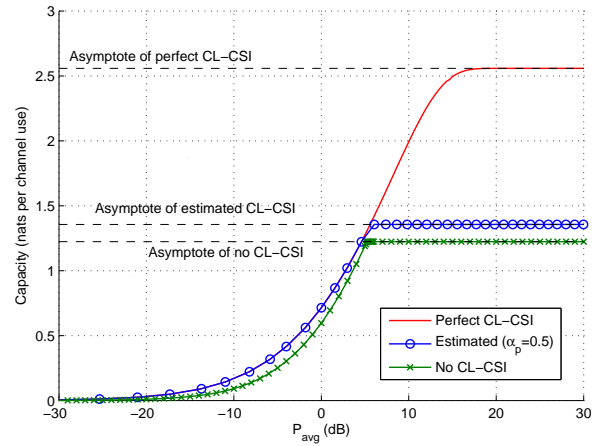

Figure 4. Ergodic capacity asymptotes for various CSI-T knowledge with $I_{peak} = 10$ dB, and $\varepsilon = 0.05$.

Fig. 4 shows ergodic capacity versus the average transmit power with $\alpha_p \in \{0, 0.5, 1\}$. We notice that, except the perfect

CL-CSI capacity curve, capacities curves for $\alpha_p = 1$ and $\alpha_p = 0.5$, both reach the asymptotes above a specific threshold $P^*_{avg}$. An important remark is that if the average power is greater than $P^*_{avg}$ given in (14), the capacity is not increasing and the supplementary power is considered as a wasted cost. Consequently, when designing a spectrum sharing system, this average power threshold should be taken into consideration in order to avoid high power consumption with no capacity return.

*3) Effect of the Interference Peak $I_{peak}$:* The interference peak, $I_{peak}$, presents the maximal interference level tolerated by the PU when the SU wants to use the same spectrum. Hence, $I_{peak}$ is an important parameter that determines the capacity limit at high SNR. The plot of the capacity as a function of $I_{peak}$ is presented in Fig. 5.

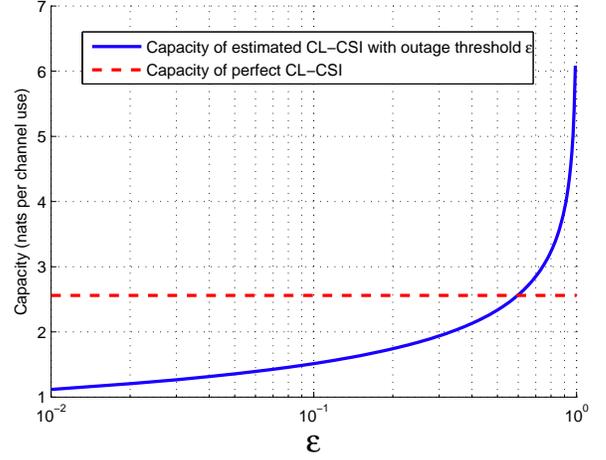

Figure 6. Outage threshold effect on the capacity at high SNR with $P_{avg} = 20$ dB, $I_{peak} = 10$ dB, and $\alpha_p = 0.5$.

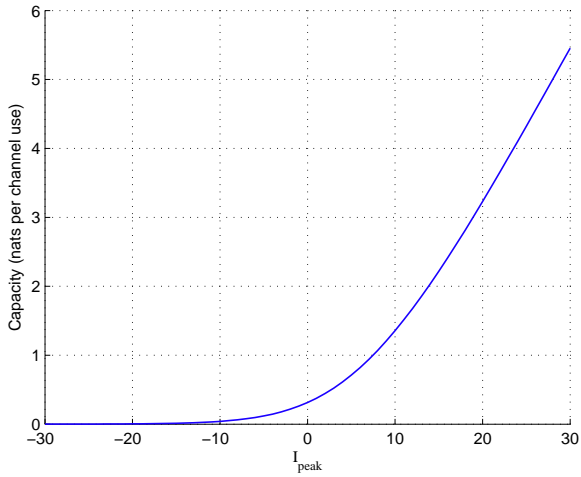

Figure 5. Interference peak effect on the capacity at high SNR with $P_{avg} = 20$ dB.

If $I_{peak}$ is high, it means that more interference is allowed and, hence, the capacity limit increases to reach the single user capacity. In fact, allowing a high level of interference (i.e. $I_{peak} \to \infty$) means that the interference constraint has no effect and the problem becomes the same as in a single user scenario.

*4) Effect of the Outage Threshold $\varepsilon$:* The outage threshold $\varepsilon$ is also a decisive parameter that affects the capacity limit at high SNR when the CL-CSI is not known. As the interference constraint is dynamic and depends on the fluctuation of the CL-CSI, $\varepsilon$ reflects the tolerated overcome of the interference peak by the secondary signal interference. The variation of the capacity at high SNR as a function of $\varepsilon$ is presented in Fig. 6.

As expected the capacity limit increases as $\varepsilon$ does and can even overcome the perfectly known CL-CSI capacity. In fact, increasing $\varepsilon$ means that the interference constraint is given less importance and the probability that the power will be below the interference threshold is low. Consequently, the capacity increases as the primary user is more tolerant toward overpassing the interference constraint. In practical communication, Fig. 6 can be interpreted as a capacity-outage tradeoff that describes the price to pay by the SU to achieve a certain rate.

*5) Effect of the CL-CSI Estimation Error $\alpha_p$:* In order to study the effect of the CL-CSI estimation variance, $\alpha_p$, on the capacity at high SNR with a fixed $P_{avg}$, the variation of the capacity with $\alpha_p$ is plotted numerically in Fig. 7 with fixed $P_{avg} = 20$ dB.

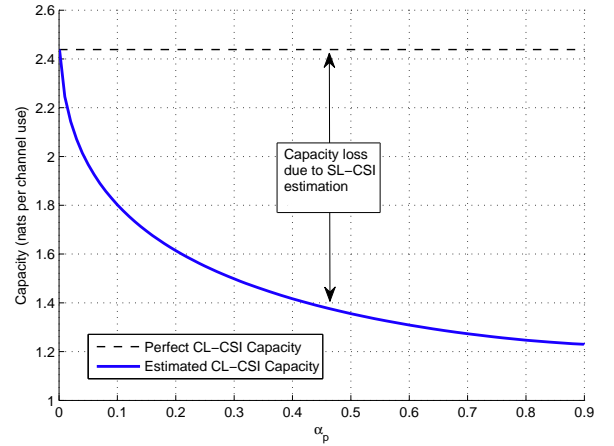

Figure 7. CL-CSI Estimation error effect on the capacity at high SNR limit with $P_{avg} = 20$ dB, $I_{peak} = 10$ dB, and $\varepsilon = 0.05$.

We notice that as the error increases between 0 and 0.3, the capacity decreases rapidly. However, when the error is high (> 0.5), the variation of the capacity is low. Consequently, when estimating the CL-CSI and the range of the error is known previously, more resources should be invested and accurate methods have to be adopted used at low range $[0, 0.3]$. In contrast, in high rang $[0.5, 1]$ accuracy is not important since the capacity gain is not that rewarding.

## VI. ON-OFF POWER SCHEME

In this section, we present an on-off power control scheme and evaluate its achievable rate. The proposed on-off power

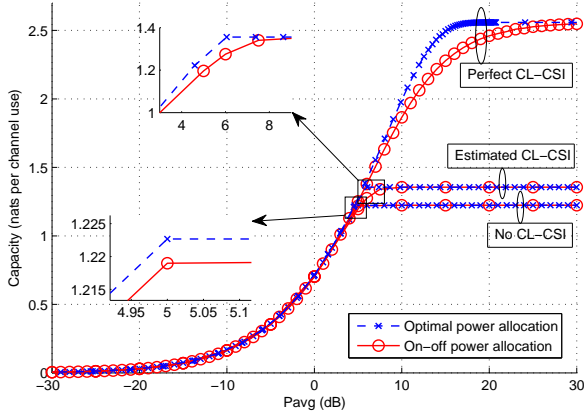

Figure 8. On-off scheme in CR with various knowledge of the CL-CSI.

scheme is given by

$$P = \begin{cases} P_0 & \text{if } |h_s|^2 \geq \tau \\ 0 & \text{otherwise,} \end{cases} \quad (34)$$

where $P_0$ is the power level in the "On" mode and $\tau$ is a non negative real presenting a threshold below which no transmission is performed. From this power profile, the value of the channel gain $h_s$ is needed in order to decide whether the power should switch on or off. Note that, this problem is less complex than the previous problems since we are optimizing in the positive real space rather than the positive continuous functions space. In the CR context, the interference constraint is introduced. Since, the peak power constraint is dynamic, $P$ should satisfy the constraint for each value of the random channel gain $h_p$. Consequently, the "On" level will be also dynamic and will be the combination of single on-off power and the CR adapted power. Meanwhile, obtaining the optimal value of $\tau$ is performed numerically by solving the following optimization problem:

$$\max \quad \int_\tau^\infty \log\left(1 + P_0 \times |h_s|^2\right) f_{|h_s|^2}(|h_s|^2) \, d|h_s|^2 \quad (35)$$

$$\text{subject to} \quad \tau \geq 0. \quad (36)$$

### A. No CL-CSI

When the CL-CSI is not known, the optimal power is, therefore, given by

$$P_0 = \min\left\{\frac{P_{avg}}{\left(1 - F_{|h_s|^2}(\tau)\right)}, \frac{I_{peak}}{F^{-1}_{|h_p|^2}(1-\varepsilon)}\right\}. \quad (37)$$

The simulation in Rayleigh fading, given in Fig. 8, shows that the on-off rate is close to the channel capacity for all SNR values except the mid-range SNR (4 - 8 dB) and we notice that

- if $P_{avg} < \frac{I_{peak}}{F^{-1}_{|h_p|^2}(1-\varepsilon)}$, the optimal threshold $\tau$ is the same as for the on-off single user scheme
- if $P_{avg} \geq \frac{I_{peak}}{F^{-1}_{|h_p|^2}(1-\varepsilon)}$, $\tau = 0$ and $P_0 = \frac{I_{peak}}{F^{-1}_{|h_p|^2}(1-\varepsilon)}$.

### B. Perfect CL-CSI

When the CL-CSI is perfectly known at the SU_Tx, the power profile is given by

$$P_0 = \min\left\{\frac{P_{avg}}{\left(1 - F_{|h_s|^2}(\tau)\right)}, \frac{I_{peak}}{|h_p|^2}\right\}. \quad (38)$$

Considering this power profile, the resulting capacity is found by solving the following maximization problem

$$\max_{\tau \geq 0} \quad \int_0^\infty \left(\int_\tau^\infty \log\left(1 + \min\left\{\frac{P_{avg}}{\left(1 - F_{|h_s|^2}(\tau)\right)}, \frac{I_{peak}}{\nu}\right\} \gamma\right) f_{|h_s|^2}(\gamma) \, d\gamma\right) f_{|h_p|^2}(\nu) \, d\nu \quad (39)$$

$$\text{subject to} \quad \tau \geq 0. \quad (40)$$

The corresponding CR on-off rate $R$ in Rayleigh fading is given by

$$R = \max_{\tau \geq 0} \int_0^\infty \left(\int_\tau^\infty \log\left(1 + \min\left\{\frac{P_{avg}}{e^{-\tau}}, \frac{I_{peak}}{\nu}\right\} \gamma\right) e^{-\gamma} \, d\gamma\right) e^{-\nu} \, d\nu. \quad (41)$$

Solving the problem given in (41) is performed numerically and the results are presented in Fig. 8. We notice that the on-off rate is, again, matching the actual capacity at low and high SNR. In mid-range SNR (5 - 25 dB), the on-off rate is relatively far from the actual capacity. This gap is due to the fact that in this region, both power and interference constraints are involved including the fixed power (in the "On" mode) which does not achieve the capacity.

### C. Estimated CL-CSI

With estimated CL-CSI, the power profile is given by

$$P_0 = \min\left\{\frac{P_{avg}}{\left(1 - F_{|h_s|^2}(\tau)\right)}, \frac{I_{peak}}{F^{-1}_{|h_p|^2|\hat{h}_p}(1-\varepsilon, \hat{h}_p)}\right\}. \quad (42)$$

The corresponding capacity is found by solving the following maximization problem

$$\max \quad \int_{\hat{h}_p} \left(\int_\tau^\infty \log\left(1 + \min\left\{\frac{P_{avg}}{\left(1 - F_{|h_s|^2}(\tau)\right)}, \frac{I_{peak}}{F^{-1}_{|h_p|^2|\hat{h}_p}(1-\varepsilon, \delta)}\right\} \gamma\right) \times f_{|h_s|^2}(\gamma) \, d\gamma\right) f_{\hat{h}_p}(\delta) \, d\delta \quad (43)$$

$$\text{subject to} \quad \tau \geq 0. \quad (44)$$

The previous maximization problem capacity is solved with Rayleigh fading distributions and results are given in Fig. 8. The chosen estimation error is $\alpha_p = 0.5$ with fixed $I_{peak} = 10$ dB and $\varepsilon = 0.05$. Again, the on-off scheme achieves a near-optimal performance at high and low SNR regimes, whereas in mid-range SNR (4 - 8 dB), a small gap between the achievable rate and the capacity is showing up.

## VII. Conclusion

In this paper, we have studied a spectrum sharing scenario with various CSI estimations at the SU_Tx both cross link and secondary link under average transmit power and interference outage constraints. We have derived a general expression for the optimal power allocation and the corresponding ergodic capacity. The obtained results represent a unified expression of the power allocation for different combinations of secondary and cross links CSI knowledge at the SU_Tx, under these constraints. To validate this general result, we have applied the unified capacity expression on previously studied scenarios and have obtained identical results that confirms our findings. We have also applied our unified framework to easily deduce the optimal power and the capacity expressions in a non-trivial scenario that has not been studied previously. On another hand, we have presented an asymptotic analysis of the capacity at low and high SNR. It has been shown that at low SNR, the capacity only depends on the knowledge of the SL-CSI. Whereas, at high SNR, the capacity has horizontal asymptotes that are constrained by the knowledge of the CL-CSI and the interference parameters. In order to show the effect of these parameters, various numerical results have been presented.

Finally, an on-off scheme has been introduced as a practical power control scheme in order to achieve part of the capacity. Adopting this scheme when both CSI are not available has led to an achievable rate that is very close to the actual capacity results at low and high SNR regimes and presents a gap in the mid-range SNR.

## Appendix I : Proof of the inequality (7)

From (6), we have:

$$\text{Prob}\left\{P|h_p|^2 \geq I_{peak}|\hat{h}_s, \hat{h}_p\right\} \leq \varepsilon$$
$$\iff \text{Prob}\left\{|h_p|^2 \geq \frac{I_{peak}}{P}|\hat{h}_s, \hat{h}_p\right\} \leq \varepsilon$$
$$\iff \int_{\frac{I_{peak}}{P}}^{\infty} f_{|h_p|^2|\hat{h}_p}(\gamma, \hat{h}_p)\, d\gamma \leq \varepsilon$$
$$\iff 1 - F_{|h_p|^2|\hat{h}_p}\left(\frac{I_{peak}}{P}, \hat{h}_p\right) \leq \varepsilon$$
$$\iff 1 - \varepsilon \leq F_{|h_p|^2|\hat{h}_p}\left(\frac{I_{peak}}{P}, \hat{h}_p\right)$$
$$\iff F^{-1}_{h_p|\hat{h}_p}(1-\varepsilon, \hat{h}_p) \leq \frac{I_{peak}}{P}$$
$$\iff P \leq \frac{I_{peak}}{F^{-1}_{|h_p|^2|\hat{h}_p}(1-\varepsilon, \hat{h}_p)}.$$

## Appendix II : Proof of the Theorem

We solve the problem by considering two optimization sub-problems, in which we have the objective function and one of the constraints. After finding the optimal power of each sub-problem, we pick the intersection between the two solution spaces, i.e. their minimum.

Now, we formulate the Lagrangian of the problem, $\mathcal{L}$, and we compute $P(\hat{h}_s, \hat{h}_p)$ such as $\frac{\partial \mathcal{L}}{\partial P(\hat{h}_s, \hat{h}_p)} = 0$ taking all realizations of $\hat{h}_s$ and $\hat{h}_p$. We obtain:

$$\frac{\partial \mathcal{L}}{\partial P(\hat{h}_s, \hat{h}_p)} = \mathbb{E}_{h_s}\left[\frac{|h_s|^2}{1 + P(\hat{h}_s, \hat{h}_p) \times |h_s|^2} - (\lambda + \mu)\right] = 0.$$

The resulting solution is given by (11) presenting a necessary and sufficient optimality condition.

In order to solve this problem, we should find optimal values of the Lagrange multipliers, noted $\lambda^*$ and $\mu^*$.

**1) Finding $\mu^*$:** The corresponding KKT complementary slackness condition [30] is given by

$$\mu \times (P - P_{I_{peak},\varepsilon}(\hat{h}_p)) = 0. \quad (45)$$

Let $I^{-1}_{\hat{h}_s}(\cdot)$ is the integral function defined by

$$I_{\hat{h}_s}(P) = \int_0^{\infty} \frac{\gamma}{1 + P\gamma} f_{|h_s|^2|\hat{h}_s}(\gamma, \hat{h}_s)\, d\gamma. \quad (46)$$

With respect to the constraint (6), we distinguish two cases:
- $P < P_{I_{peak},\varepsilon}(\hat{h}_p) \Longrightarrow \mu^* = 0$ and P is given by

$$P(\hat{h}_s, \hat{h}_p) = \left[I^{-1}_{\hat{h}_s}(\lambda)\right]^+.$$

- $P = P_{I_{peak},\varepsilon}(\hat{h}_p) \Longrightarrow \mu^* \geq 0$.

These two cases could be seen as

$$P = \min\left\{\left[I^{-1}_{\hat{h}_s}(\lambda)\right]^+, P_{I_{peak},\varepsilon}(\hat{h}_p)\right\}.$$

**2) Finding $\lambda^*$:** The corresponding KKT complementary slackness condition is given by

$$\lambda \times \left(\mathbb{E}_{\hat{h}_s, \hat{h}_p}[P] - P_{avg}\right) = 0. \quad (47)$$

With respect to the constraint (5) and the resulting power, after finding $\mu^*$, we distinguish two cases:
- $\mathbb{E}_{\hat{h}_s, \hat{h}_p}[P] < P_{avg} \Longrightarrow \lambda^* = 0$.
- $\mathbb{E}_{\hat{h}_s, \hat{h}_p}[P] = P_{avg} \Longrightarrow \lambda^* \geq 0$ and is computed by solving the following equality

$$\mathbb{E}_{\hat{h}_s, \hat{h}_p}\left[\min\left\{\left[I^{-1}_{\hat{h}_s}(\lambda)\right]^+, P_{I_{peak},\varepsilon}(\hat{h}_p)\right\}\right] = P_{avg}. \quad (48)$$

To conclude, the optimal power profile is given by

$$P(\hat{h}_s, \hat{h}_p) = \min\left\{\left[I^{-1}_{\hat{h}_s}(\lambda)\right]^+, P_{I_{peak},\varepsilon}(\hat{h}_p)\right\}, \quad (49)$$

where $\lambda$ is a solution of (48).

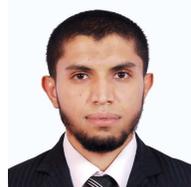

**Lokman Sboui** (S'11) was born in Cairo, Egypt. He received the Diplôme d'Ingénieur degree from École Polytechnique de Tunisie (EPT), La Marsa, Tunisia, in 2011. He is currently a Master student in the Electrical Engineering program of King Abdullah University of Science and Technology (KAUST), Thuwal, Makkah Province, Saudi Arabia. His current research interests include performance of cognitive radio systems, low SNR communication, and MIMO communication systems.

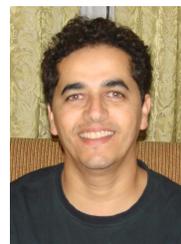

**Zouheir Rezki** (S'01, M'08) was born in Casablanca, Morocco. He received the Diplôme d'Ingénieur degree from the École Nationale de l'Industrie Minérale (ENIM), Rabat, Morocco, in 1994, the M.Eng. degree from École de Technologie Supérieure, Montreal, Québec, Canada, in 2003, and the Ph.D. degree from École Polytechnique, Montreal, Québec, in 2008, all in electrical engineering. From October 2008 to September 2009, he was a postdoctoral research fellow with Data Communications Group, Department of Electrical and Computer Engineering, University of British Columbia. He is now a research scientist at King Abdullah University of Science and Technology (KAUST), Thuwal, Mekkah Province, Saudi Arabia. His research interests include: performance limits of communication systems, cognitive and sensor networks, physical-layer security, and low-complexity detection algorithms.

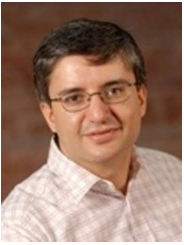

**Mohamed-Slim Alouini** (S'94, M'98, SM'03, F09) was born in Tunis, Tunisia. He received the Ph.D. degree in electrical engineering from the California Institute of Technology (Caltech), Pasadena, CA, USA, in 1998. He was with the department of Electrical and Computer Engineering of the University of Minnesota, Minneapolis, MN, USA, then with the Electrical and Computer Engineering Program at the Texas A &M University at Qatar, Education City, Doha, Qatar. Since June 2009, he has been a Professor of Electrical Engineering at King Abdullah University of Science and Technology (KAUST), Makkah Province, Saudi Arabia, where his current research interests include the design and performance analysis of wireless communication systems.

Table I
OPTIMAL POWER COMPONENTS DEPENDING ON THE AVAILABLE CSI

| Power (Scenario) | No CSI | Perfect CSI | Estimated CSI |
|---|---|---|---|
| $P_{P_{avg}}(h_s)$ (SL-CSI) | $P_{avg}$ | $\left[\frac{1}{\lambda} - \frac{1}{|h_s|^2}\right]^+$ | $\left[I_{\hat{h}_s}^{-1}(\lambda)\right]^+$ |
| $P_{I_{peak}}(h_p)$ (CL-CSI) | $\frac{I_{peak}}{F_{|h_p|^2}^{-1}(1-\varepsilon)}$ | $\frac{I_{peak}}{|h_p|^2}$ | $\frac{I_{peak}}{F_{|h_p|^2|\hat{h}_p}^{-1}(1-\varepsilon,\hat{h}_p)}$ |

where $I_{\hat{h}_s}(P)$ is defined as $I_{\hat{h}_s}(P) = \int_0^\infty \frac{\gamma}{1+P\gamma} f_{h_s|\hat{h}_s}(\gamma, \hat{h}_s)\, d\gamma$.

Table II
ELEMENTS OF $H_0$ AND $H_1$ DEPENDING ON AVAILABLE CSI

| | CL-CSI \ SL-CSI | No SL-CSI | Perfect SL-CSI | Estimated SL-CSI |
|---|---|---|---|---|
| $H_0$ | No CL-CSI | The power profile | $h_s$ | $\hat{h}_s$ |
| | Perfect CL-CSI | does not depend | $h_s, h_p$ | $\hat{h}_s, h_p$ |
| | Estimated CL-CSI | on $\lambda$ | $h_s, h_p$ | $h_s, h_p$ |
| $H_1$ | No CSI | $h_s$ | $h_s$ | $h_s, \hat{h}_s$ |
| | Perfect CSI | $h_s, h_p$ | $h_s, h_p$ | $h_s, \hat{h}_s, h_p$ |
| | Estimated CSI | $h_s, \hat{h}_p$ | $h_s, \hat{h}_p$ | $h_s, \hat{h}_s, \hat{h}_p$ |

Table III
ERGODIC CAPACITY LIMITS AT LOW SNR

| Scenario | General Capacity | Capacity Asymptotes over Rayleigh Fading Channels |
|---|---|---|
| No SL-CSI | $\int_{h_s} \log(1 + P_{avg}\, \gamma)\, f_{h_s}(\gamma)\, d\gamma$ | $e^{\frac{1}{P_{avg}}} E_1(\frac{1}{P_{avg}})$ [30] |
| Perfect SL-CSI | $\int_{h_s} \log(\frac{\gamma}{\lambda}) f_{h_s}(\gamma)\, d\gamma$ [30] | $E_1(g_1)$ |
| Estimated SL-CSI | $\int_{\hat{h}_s}\left(\int_{|h_s|^2}\log\left(1 + \left[I_{\hat{h}_s}^{-1}(\lambda)\right]^+ \gamma\right) f_{|h_s|^2|\hat{h}_s}(\gamma, \delta))\, d\gamma\right) f_{\hat{h}_s}(\delta)\, d\delta$ | $\left(\int_0^\infty \left(\int_0^\infty (\log(1+\left[I_{\hat{h}_s}^{-1}(\lambda)\right]^+ \gamma) \frac{1}{\alpha_s} e^{-\frac{\gamma+\delta}{\alpha_s}} I_0(\frac{2}{\alpha_s}\sqrt{\delta\gamma})\, d\gamma\right) \frac{e^{-\frac{\delta}{1-\alpha_s}}}{1-\alpha_s}\, d\delta$ |

where $E_1(\cdot)$ is the exponential integral function.

Table IV
ERGODIC CAPACITY LIMITS AT HIGH SNR

| Scenario | General Capacity | Capacity Asymptotes over Rayleigh Fading Channels |
|---|---|---|
| No CL-CSI | $\int_{|h_s|^2} \log(1 + \frac{I_{peak}}{F_{|h_p|^2}^{-1}(1-\varepsilon)}\, \gamma)\, f_{|h_s|^2}(\gamma)\, d\gamma$ | $e^{-\frac{\log(\varepsilon)}{P_{I_{peak}}(h_p)}} E_1(-\frac{\log(\varepsilon)}{P_{I_{peak}}(h_p)})$ |
| Perfect CL-CSI | $\int_{|h_p|^2}\left(\int_{|h_s|^2} \log(1 + \frac{I_{peak}}{\delta}\, \gamma) f_{|h_s|^2}(\gamma)\, d\gamma\right) f_{|h_p|^2}(\delta)\, d\delta$ [18] | $\frac{I_{peak} \log(I_{peak})}{I_{peak}-1}$ |
| Estimated CL-CSI | $\int_{\hat{h}_s}\left(\int_{|h_s|^2} \log\left(1 + \frac{I_{peak}}{F_{|h_p|^2|\hat{h}_p}^{-1}(1-\varepsilon,\delta)}\, \gamma\right) f_{|h_s|^2}(\gamma)\, d\gamma\right) f_{\hat{h}_p}(\delta)\, d\delta$ | $\int_0^\infty \left(\int_0^\infty \log(1 + \frac{I_{peak}}{F_{|h_p|^2|\hat{h}_p}^{-1}(1-\varepsilon,\delta)}\, \gamma) e^{-\gamma}\, d\gamma\right) \frac{e^{-\frac{\delta}{1-\alpha_p}}}{1-\alpha_p}\, d\delta$ [18] |